\documentclass{article}

    \PassOptionsToPackage{numbers, compress}{natbib}

    \usepackage[final]{neurips_2023}

\usepackage[utf8]{inputenc} 
\usepackage[T1]{fontenc}    
\usepackage{hyperref}       
\usepackage{url}            
\usepackage{booktabs}       
\usepackage{amsfonts}       
\usepackage{nicefrac}       
\usepackage{microtype}      
\usepackage{xcolor}         
\usepackage{graphicx}
\usepackage{placeins}
\usepackage{listings}
\usepackage{float}
\usepackage{longtable}
\usepackage{array}
\hypersetup{
  colorlinks=true,
  citecolor=teal,
  filecolor=teal,
  linkcolor=teal,
  urlcolor=teal,
  breaklinks=true,
  hypertexnames=false
}

\title{What Do I Hear? \\ Generating Sounds for Visuals with ChatGPT}

\author{
    David Chuan-En Lin \\
    Carnegie Mellon University\\
    \texttt{chuanenl@cs.cmu.edu} \\
    \And
    Nikolas Martelaro \\
    Carnegie Mellon University\\
    \texttt{nikmart@cmu.edu} \\
}

\begin{document}

\maketitle

\vspace{-1em}

\begin{figure}[h]
  \centering
  \includegraphics[width=\linewidth]{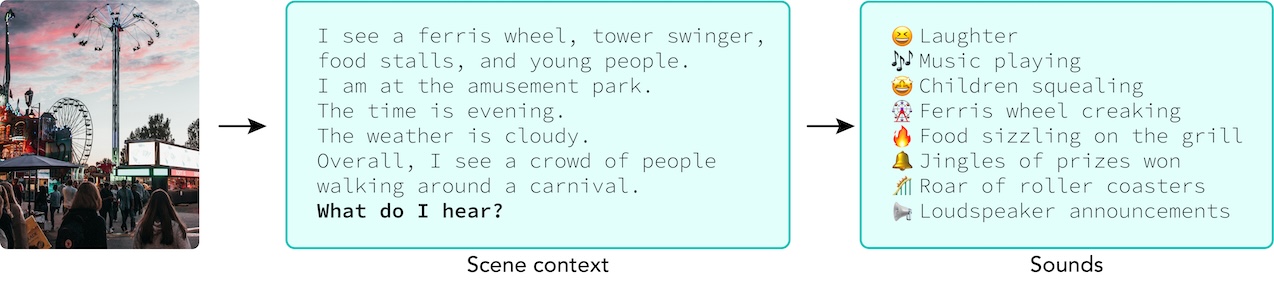}
  \caption{We first translate visual elements into a detailed scene description. We then ask ChatGPT to ideate sounds for the scene using the trigger: \textit{What do I hear?} Live demo: \url{http://soundify.cc}.}
  \label{fig:teaser}
\end{figure}

\section{Introduction}
This short paper introduces a workflow for generating realistic soundscapes for visual media. In contrast to prior work, which primarily focus on matching sounds for \textit{on-screen} visuals \cite{lin2023soundify, zhao2018sound, ghose2020autofoley}, our approach extends to suggesting sounds that \textit{may not be immediately visible} but are essential to crafting a convincing and immersive auditory environment. Our key insight is leveraging the reasoning capabilities of language models, such as ChatGPT \cite{ouyang2022training}. In this paper, we describe our workflow, which includes creating a scene context, brainstorming sounds, and generating the sounds.

\section{Method}

\subsection{Creating a Scene Context}
\begin{figure}[t]
  \centering
  \includegraphics[width=8.2cm]{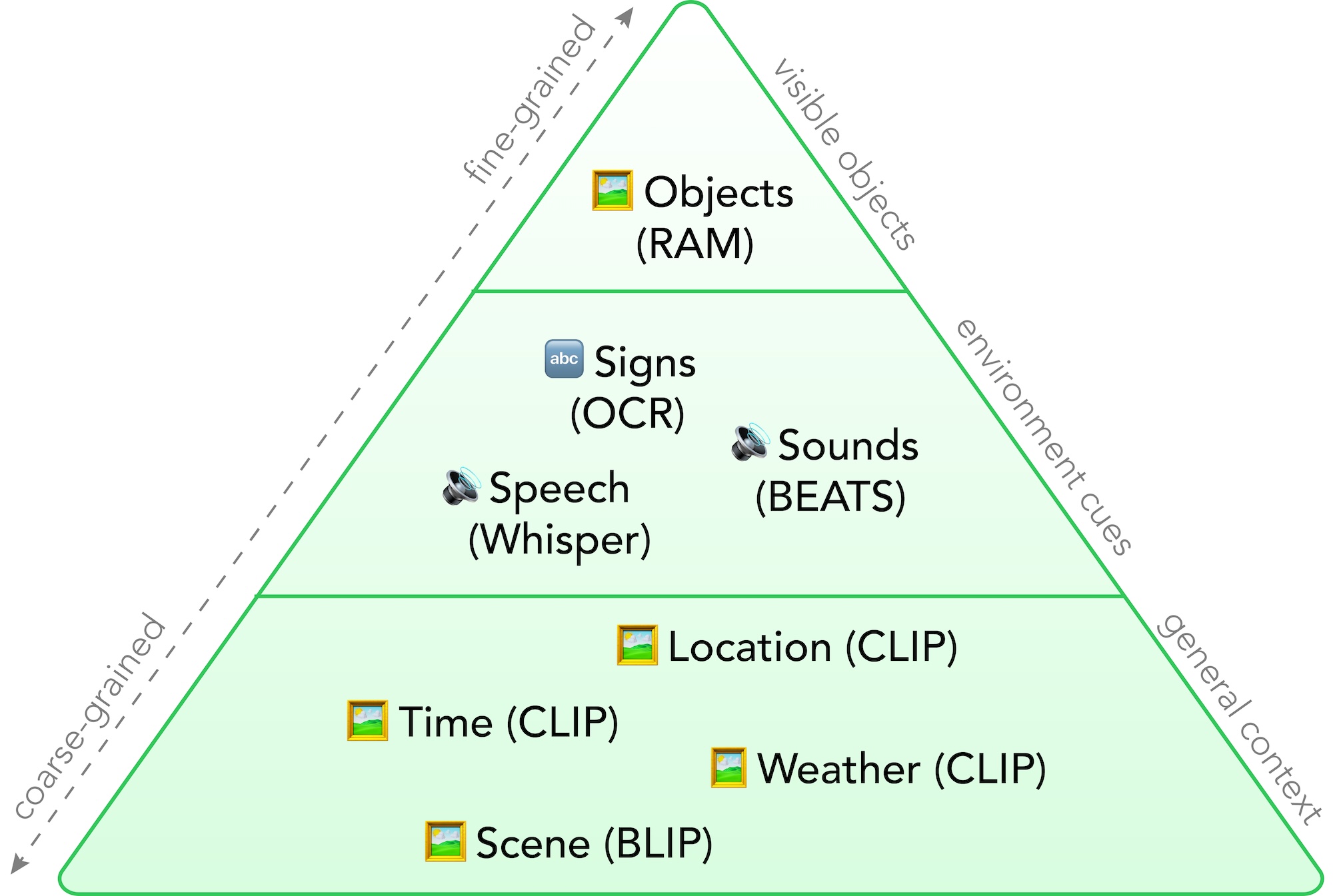}
  \caption{Soundify adapts pan (top row) and gain (bottom row) parameters over time.}
  \label{fig:scene-context}
  \vspace{-0.9em}
\end{figure}

In order to suggest realistic and fine-grained sounds for a scene, we first develop a detailed understanding of the scene. We perform scene understanding in a multi-level and multimodal manner (Figure \ref{fig:scene-context}).
The first layer focuses on the \textit{visible objects}. We use Recognize Anything \cite{zhang2023recognize} to detect object categories from a combination of the Objects365 dataset \cite{shao2019objects365} and the Open Images dataset \cite{kuznetsova2020open}.
The second layer focuses on the \textit{environment cues}, which includes text from signs, speech, and sounds. We use Tesseract OCR \cite{smith2007overview} to extract text from any visible signs, which may provide cues such as names of items, locations, or events. We use Whisper ASR \cite{radford2023robust} to transcribe any spoken words and we use BEATs \cite{chen2022beats} to identify sounds, if any.
The third layer focuses on the \textit{general context}, which includes location, time, weather, and a descriptive caption of the overall scene. We first use CLIP \cite{radford2021learning} to classify whether the scene is indoors or outdoors. If the scene is classified as indoors, we use CLIP to classify its location using categories from the Places365 dataset \cite{zhou2017places}. If the scene is classified as outdoors, in addition to classifying its location, we use CLIP to classify its time among the categories of morning, afternoon, evening, and night, and to classify its weather among the categories of sunny, foggy, windy, cloudy, thunderstorm, rainy, drizzle, snowy, and blizzard. We then use BLIP \cite{li2022blip} to generate a sentence-long descriptive caption of the overall scene.
Finally, we format the various scene understanding results into the following scene context prompt for ChatGPT:
\pagebreak
\begin{lstlisting}
I see <objects>. I am at <location>. The time is <time>. The weather is <weather>. There are sounds of <sounds>. There are signs writing <OCR text>. There are people saying <ASR transcript>. Overall, I see <scene caption>.
\end{lstlisting}

\subsection{Brainstorming Sounds}
Next, we pass the scene context into ChatGPT and prompt it with the question: \textit{What do I hear?} By utilizing the capabilities of ChatGPT, we can ideate fine-grained and diverse sound suggestions that collectively form the audio landscape of the described scene, including sounds that are not directly visible from the visuals. On our interface, we display the list of sound suggestions as tiled options (Figure \ref{fig:interface-1}). The sound designer can click on a sound suggestion to select it and also has the option to manually type in any custom sound descriptions. We assigned an emoji to each sound suggestion by prompting ChatGPT to respond with an emoji that is semantically closest. Each sound suggestion option is accompanied by a \textit{+ similar} button. When the button is clicked, we prompt ChatGPT to suggest two additional sounds that are similar to the original one for an expanded sonic palette.

\subsection{Generating the Sounds}
Once the sound designer has selected several sound suggestions, we pass each sound suggestion through AudioGen \cite{kreuk2022audiogen}, a text-to-audio model to synthesize the soundtrack. After generating the individual soundtracks for the select sound suggestions, we provide volume level predictions for each soundtrack based on the visuals. We first use the spaCy NLP library \cite{honnibal2017spacy} to extract the subject of each sound suggestion. We then use CLIP to detect whether the sound subject is visually present in the video. If the sound subject is not detected, we categorize it as a background sound and decreased the volume level by 7 dB. If the sound subject is detected, we categorize it as a foreground sound. For foreground sounds, we \textit{localize} sounds similar to Soundify \cite{lin2023soundify}. We first apply Grad-CAM \cite{selvaraju2016grad} to visualize activation maps. We then estimate the size of the subject using the activation map's area and adjust the volume level accordingly (objects that appear larger are closer in proximity and hence produce louder sounds). On the interface, we set the default values for each soundtrack according to the volume level predictions, while allowing sound designers to manually make adjustments (Figure \ref{fig:interface-2}).
We allow sound designers to listen to the combined soundtrack as they adjust the volume levels for each individual track. Finally, we provide the option for sound designers to export the final result (visuals + combined soundtracks), the combined soundtrack, or the individual soundtracks.

\section{Conclusion and Future Work}
This paper presents a workflow for generating immersive soundscapes for visual media. For future work, we would like to create an even richer scene context, such as incorporating motion cues and understanding the states of objects within the scene. In addition, we would like to investigate how to layer different sounds more effectively to achieve a harmonious blend, such as balancing frequencies.

\bibliographystyle{plainnat}
\bibliography{main}

\begin{thebibliography}{16}
\providecommand{\natexlab}[1]{#1}
\providecommand{\url}[1]{\texttt{#1}}
\expandafter\ifx\csname urlstyle\endcsname\relax
  \providecommand{\doi}[1]{doi: #1}\else
  \providecommand{\doi}{doi: \begingroup \urlstyle{rm}\Url}\fi

\bibitem[Chen et~al.(2022)Chen, Wu, Wang, Liu, Tompkins, Chen, and Wei]{chen2022beats}
Sanyuan Chen, Yu~Wu, Chengyi Wang, Shujie Liu, Daniel Tompkins, Zhuo Chen, and Furu Wei.
\newblock Beats: Audio pre-training with acoustic tokenizers.
\newblock \emph{arXiv preprint arXiv:2212.09058}, 2022.

\bibitem[Ghose and Prevost(2020)]{ghose2020autofoley}
Sanchita Ghose and John~Jeffrey Prevost.
\newblock Autofoley: Artificial synthesis of synchronized sound tracks for silent videos with deep learning.
\newblock \emph{IEEE Transactions on Multimedia}, 23:\penalty0 1895--1907, 2020.

\bibitem[Honnibal and Montani(2017)]{honnibal2017spacy}
Matthew Honnibal and Ines Montani.
\newblock spacy 2: Natural language understanding with bloom embeddings, convolutional neural networks and incremental parsing.
\newblock \emph{To appear}, 7\penalty0 (1):\penalty0 411--420, 2017.

\bibitem[Kreuk et~al.(2022)Kreuk, Synnaeve, Polyak, Singer, D{\'e}fossez, Copet, Parikh, Taigman, and Adi]{kreuk2022audiogen}
Felix Kreuk, Gabriel Synnaeve, Adam Polyak, Uriel Singer, Alexandre D{\'e}fossez, Jade Copet, Devi Parikh, Yaniv Taigman, and Yossi Adi.
\newblock Audiogen: Textually guided audio generation.
\newblock \emph{arXiv preprint arXiv:2209.15352}, 2022.

\bibitem[Kuznetsova et~al.(2020)Kuznetsova, Rom, Alldrin, Uijlings, Krasin, Pont-Tuset, Kamali, Popov, Malloci, Kolesnikov, et~al.]{kuznetsova2020open}
Alina Kuznetsova, Hassan Rom, Neil Alldrin, Jasper Uijlings, Ivan Krasin, Jordi Pont-Tuset, Shahab Kamali, Stefan Popov, Matteo Malloci, Alexander Kolesnikov, et~al.
\newblock The open images dataset v4: Unified image classification, object detection, and visual relationship detection at scale.
\newblock \emph{International Journal of Computer Vision}, 128\penalty0 (7):\penalty0 1956--1981, 2020.

\bibitem[Li et~al.(2022)Li, Li, Xiong, and Hoi]{li2022blip}
Junnan Li, Dongxu Li, Caiming Xiong, and Steven Hoi.
\newblock Blip: Bootstrapping language-image pre-training for unified vision-language understanding and generation.
\newblock In \emph{International Conference on Machine Learning}, pages 12888--12900. PMLR, 2022.

\bibitem[Lin et~al.(2023)Lin, Germanidis, Valenzuela, Shi, and Martelaro]{lin2023soundify}
David Chuan-En Lin, Anastasis Germanidis, Crist{\'o}bal Valenzuela, Yining Shi, and Nikolas Martelaro.
\newblock Soundify: Matching sound effects to video.
\newblock In \emph{Proceedings of the 36th Annual ACM Symposium on User Interface Software and Technology}, 2023.

\bibitem[Ouyang et~al.(2022)Ouyang, Wu, Jiang, Almeida, Wainwright, Mishkin, Zhang, Agarwal, Slama, Ray, et~al.]{ouyang2022training}
Long Ouyang, Jeffrey Wu, Xu~Jiang, Diogo Almeida, Carroll Wainwright, Pamela Mishkin, Chong Zhang, Sandhini Agarwal, Katarina Slama, Alex Ray, et~al.
\newblock Training language models to follow instructions with human feedback.
\newblock \emph{Advances in Neural Information Processing Systems}, 35:\penalty0 27730--27744, 2022.

\bibitem[Radford et~al.(2021)Radford, Kim, Hallacy, Ramesh, Goh, Agarwal, Sastry, Askell, Mishkin, Clark, et~al.]{radford2021learning}
Alec Radford, Jong~Wook Kim, Chris Hallacy, Aditya Ramesh, Gabriel Goh, Sandhini Agarwal, Girish Sastry, Amanda Askell, Pamela Mishkin, Jack Clark, et~al.
\newblock Learning transferable visual models from natural language supervision.
\newblock In \emph{International conference on machine learning}, pages 8748--8763. PMLR, 2021.

\bibitem[Radford et~al.(2023)Radford, Kim, Xu, Brockman, McLeavey, and Sutskever]{radford2023robust}
Alec Radford, Jong~Wook Kim, Tao Xu, Greg Brockman, Christine McLeavey, and Ilya Sutskever.
\newblock Robust speech recognition via large-scale weak supervision.
\newblock In \emph{International Conference on Machine Learning}, pages 28492--28518. PMLR, 2023.

\bibitem[Selvaraju et~al.(2016)Selvaraju, Das, Vedantam, Cogswell, Parikh, and Batra]{selvaraju2016grad}
Ramprasaath~R Selvaraju, Abhishek Das, Ramakrishna Vedantam, Michael Cogswell, Devi Parikh, and Dhruv Batra.
\newblock Grad-cam: Why did you say that?
\newblock \emph{arXiv preprint arXiv:1611.07450}, 2016.

\bibitem[Shao et~al.(2019)Shao, Li, Zhang, Peng, Yu, Zhang, Li, and Sun]{shao2019objects365}
Shuai Shao, Zeming Li, Tianyuan Zhang, Chao Peng, Gang Yu, Xiangyu Zhang, Jing Li, and Jian Sun.
\newblock Objects365: A large-scale, high-quality dataset for object detection.
\newblock In \emph{Proceedings of the IEEE/CVF international conference on computer vision}, pages 8430--8439, 2019.

\bibitem[Smith(2007)]{smith2007overview}
Ray Smith.
\newblock An overview of the tesseract ocr engine.
\newblock In \emph{Ninth international conference on document analysis and recognition (ICDAR 2007)}, volume~2, pages 629--633. IEEE, 2007.

\bibitem[Zhang et~al.(2023)Zhang, Huang, Ma, Li, Luo, Xie, Qin, Luo, Li, Liu, et~al.]{zhang2023recognize}
Youcai Zhang, Xinyu Huang, Jinyu Ma, Zhaoyang Li, Zhaochuan Luo, Yanchun Xie, Yuzhuo Qin, Tong Luo, Yaqian Li, Shilong Liu, et~al.
\newblock Recognize anything: A strong image tagging model.
\newblock \emph{arXiv preprint arXiv:2306.03514}, 2023.

\bibitem[Zhao et~al.(2018)Zhao, Gan, Rouditchenko, Vondrick, McDermott, and Torralba]{zhao2018sound}
Hang Zhao, Chuang Gan, Andrew Rouditchenko, Carl Vondrick, Josh McDermott, and Antonio Torralba.
\newblock The sound of pixels.
\newblock In \emph{Proceedings of the European conference on computer vision (ECCV)}, pages 570--586, 2018.

\bibitem[Zhou et~al.(2017)Zhou, Lapedriza, Khosla, Oliva, and Torralba]{zhou2017places}
Bolei Zhou, Agata Lapedriza, Aditya Khosla, Aude Oliva, and Antonio Torralba.
\newblock Places: A 10 million image database for scene recognition.
\newblock \emph{IEEE transactions on pattern analysis and machine intelligence}, 40\penalty0 (6):\penalty0 1452--1464, 2017.

\end{thebibliography}
\raggedbottom
\pagebreak

\appendix

\section{Additional Examples}

\begin{longtable}{|m{0.2\columnwidth}|m{0.2\columnwidth}|m{0.6\columnwidth}|}
\hline
\textbf{Scene Description} & \textbf{Visual} & \textbf{Predicted Sounds} \\ \hline

Cafe &
  \includegraphics[width=0.2\columnwidth]{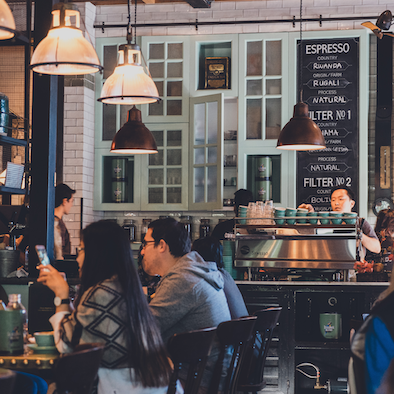} &
  Clinking of silverware, Murmuring of conversations, Hum of espresso machine, Clack of cash register, Jingle of doorbell \\ \hline
Park &
  \includegraphics[width=0.2\columnwidth]{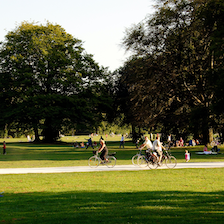} &
  Birds chirping, Dogs barking, Children laughing, Wind rustling through trees, Leaves crunching underfoot \\ \hline
Library &
  \includegraphics[width=0.2\columnwidth]{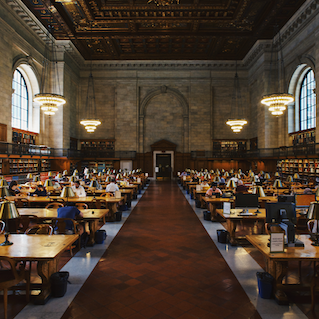} &
  Pages rustling, Pencils scratching, Quiet conversations, Keyboard tapping, Soft footsteps \\ \hline
Train Station &
  \includegraphics[width=0.2\columnwidth]{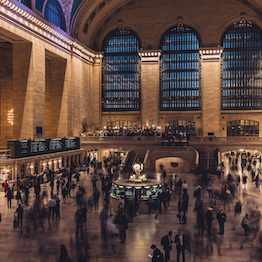} &
  Ticket machines beeping, Footsteps on concrete, Trains arriving and departing, Children laughing and shouting, Murmurs of conversation \\ \hline
Concert &
  \includegraphics[width=0.2\columnwidth]{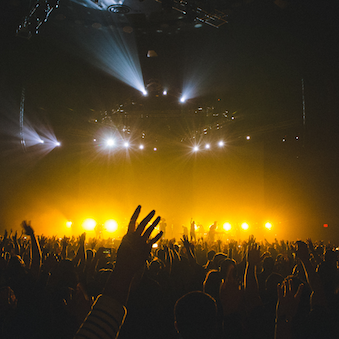} &
  Cheering, Music blaring, Bass thumping, Clapping hands, Vocals singing \\ \hline
Airport &
  \includegraphics[width=0.2\columnwidth]{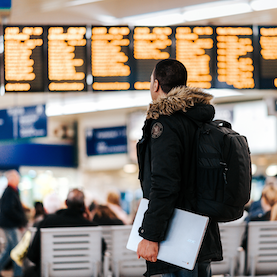} &
  Announcements over the intercom, Rolling of suitcases on the floor, Airplanes taking off and landing, Beeping of scanners at security checkpoints, Whirring of air conditioning units \\ \hline
Rainforest &
  \includegraphics[width=0.2\columnwidth]{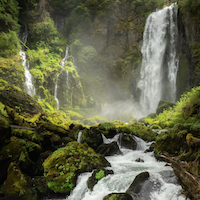} &
  Echoing cascades, Rustling leaves, Trickling water, Chirping birds, Distant thunder
  \\ \hline

\caption{Examples scenes and predicted sounds} \label{tab:scenes_sounds}
\end{longtable}

\pagebreak

\begin{longtable}{|p{0.2\columnwidth}|m{0.2\columnwidth}|m{0.6\columnwidth}|}
\hline
\textbf{Scene Description} & \textbf{Visual} & \textbf{Predicted Sounds} \\ \hline

Moon -- Silent Scene &
  \includegraphics[width=0.2\columnwidth]{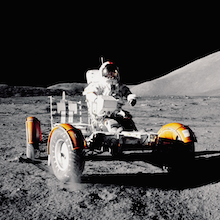} &
  Occasional crackle of radio static, Quiet whirring of machinery, Gentle pattering of sand, Clanking of tools being used by the astronaut, Buzzing of a nearby satellite \\ \hline
Cyberpunk City -- AI-generated Scene &
  \includegraphics[width=0.2\columnwidth]{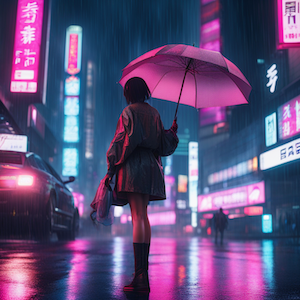} &
  Gentle drumming of raindrops on umbrella, Splashing of puddles, Hum of traffic, Sirens wailing, People shouting in the distance \\ \hline
The Legend of Zelda -- Video Game Scene &
  \includegraphics[width=0.2\columnwidth]{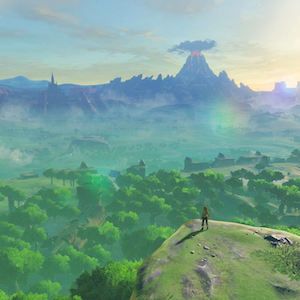} &
  Birds chirping, Wind rustling through the trees, A stream trickling nearby, A farmer whistling a tune, Cows mooing in the distance \\ \hline
\caption{Stress testing examples} \label{tab:scenes_sounds}
\end{longtable}

\pagebreak

\section{User Interface}

\begin{figure}[H]
  \centering
  \includegraphics[width=12cm]{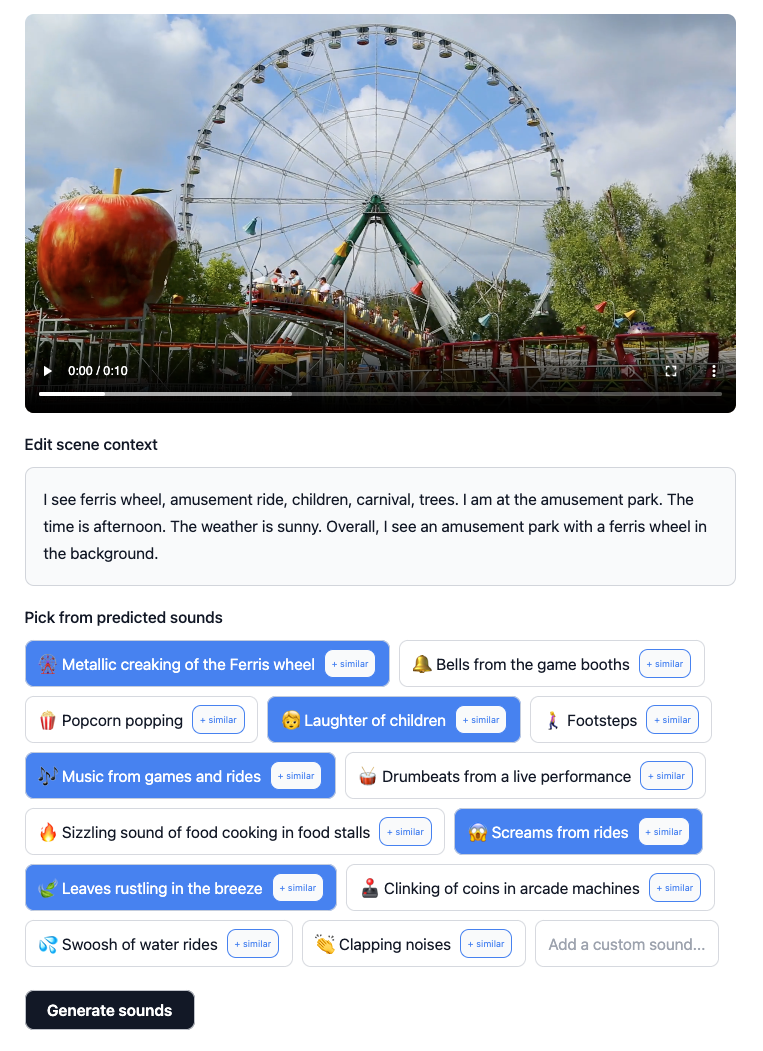}
  \caption{Using the interface, the user can select the sounds to generate, add any custom sound descriptions, and ideate additional similar sounds.}
  \label{fig:interface-1}
\end{figure}

\begin{figure}[H]
  \centering
  \includegraphics[width=12cm]{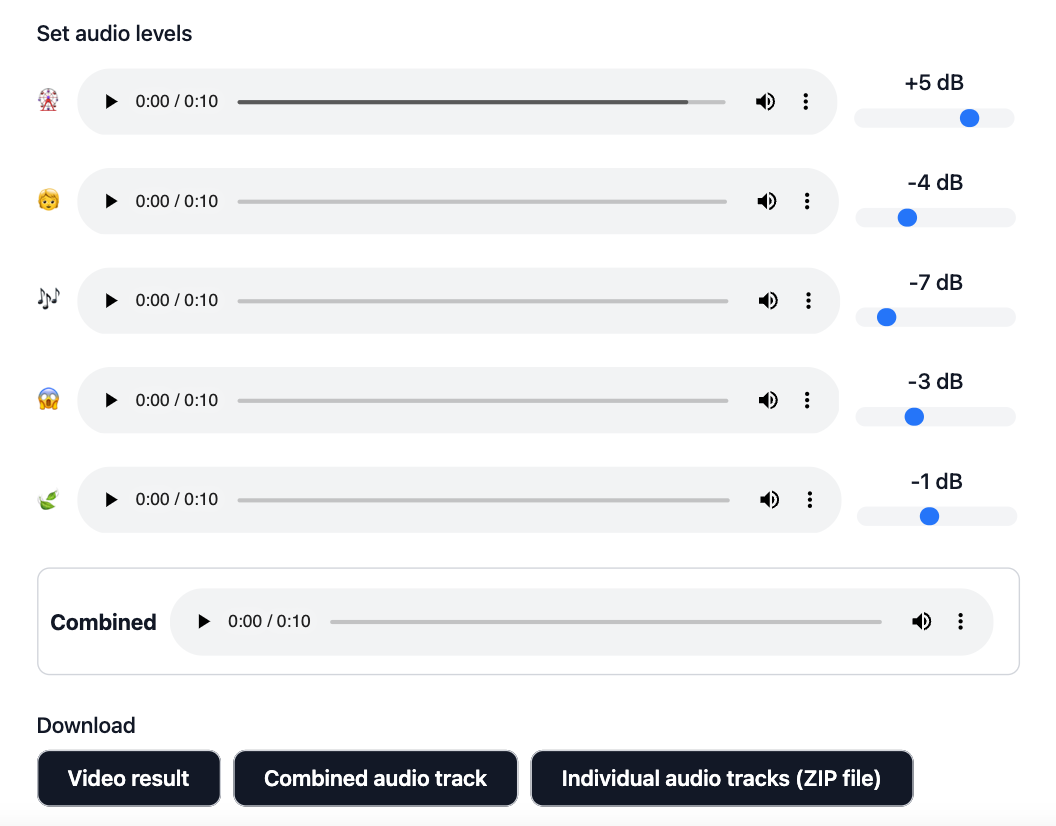}
  \caption{Using the interface, the user can edit the volume levels of each soundtrack, listen to the combined soundtrack, and export the soundtracks.}
  \label{fig:interface-2}
\end{figure}

\end{document}